# Entangled Causal Relation in Measurement by Two Measurement Devices Moving Mutually

Narumi Ohkawa

*Department of Device Integration, Fujitsu Mie Plant, Kuwana 511-0192*

In this paper, I want to show situations in which causal relation of two events is entangled and similar to time paradox, without employing time machines. Such situations will be obtained in measurement by two measurement devices moving mutually. According to the concept of the collapse of wave packet and the relativity of simultaneity, the causal relation of two measurements can be entangled.

## 1. Introduction

Recently time machine is discussed as an issue of physics [1-2]. And time paradox or causality paradox is also an attractive issue of physics [3-4]. For example, one of time paradoxes is a situation in which the cause and the result can not be distinguished between two events in causal relation. It seems not easy to realize time machine and time paradox by current technology. But, situations similar to time paradox can be realized without employing time machine. According to the concept of the collapse of wave packet and the relativity of simultaneity, situations in which causal relation of two events is entangled will be obtained in measurement by two measurement devices moving mutually.

## 2. Collapse of wave packet and relativity of simultaneity

We consider the case that one electron is split and advance toward left and right, and two measurements are done at both sides. We call these measurements measurement (L) and measurement (R). Now suppose that two measurements are done by two measurement devices which are moving mutually as shown in Fig. 1. Two measurement devices are going away from each other or approaching to each other with velocity V and -V. According to the relativity of simultaneity, it is relative whether two events at different points are simultaneous or not. So it is important when and in which coordinate system the collapse of wave packet occurs. It seems natural to think as follows.

Hypothesis: the collapse of wave packet occurs at the time when the measurement is done with respect to the coordinate system in which the measurement device is at rest.

We call such coordinate systems for above two measurements L and R. Once we accept this hypothesis, situations in which causal relation is entangled are obtained.

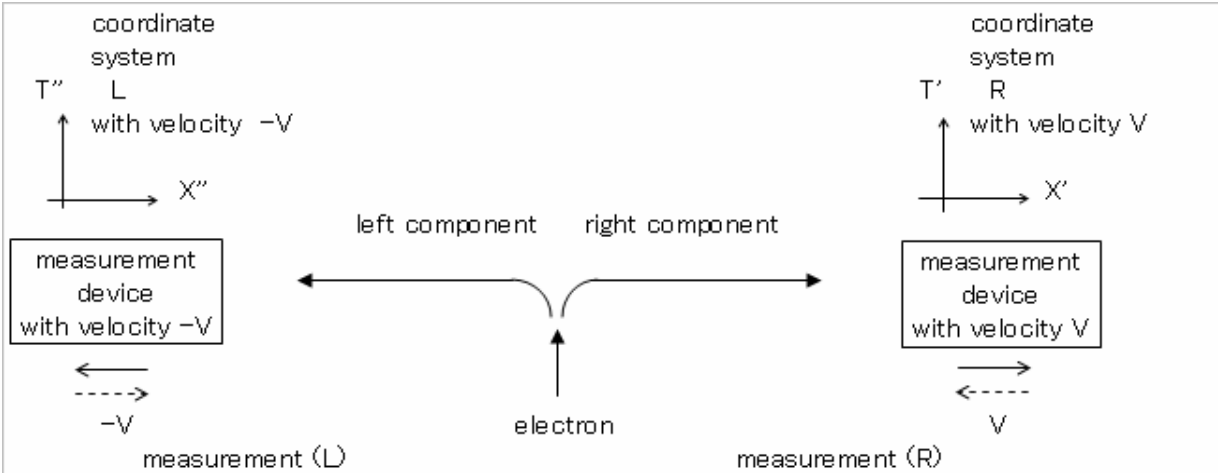

Fig. 1.  Measurement of split electron by two measurement devices moving mutually. X'-T' is the coordinate system R of measurement (R) moving with the velocity V. X"-T" is the coordinate system L of measurement (L) moving with the velocity -V.

### 3. Case1: A situation similar to one of time paradoxes

We consider the case in which two measurement devices are going away from each other. As shown in Fig. 2, according to the relativity of simultaneity, measurement (R) is done before measurement (L) with respect to R. And measurement (L) is done before measurement (R) with respect to L. According to above hypothesis, each measurement can cause the collapse of wave packet. But this also means that each measurement is done after the collapse of wave packet caused by the other measurement. So each measurement can not cause the collapse of wave packet. This situation is similar to one of time paradoxes because we can not know which measurement is the cause and which measurement is the result. I believe it is very interesting to confirm what actually happens in the situation by the experiment.

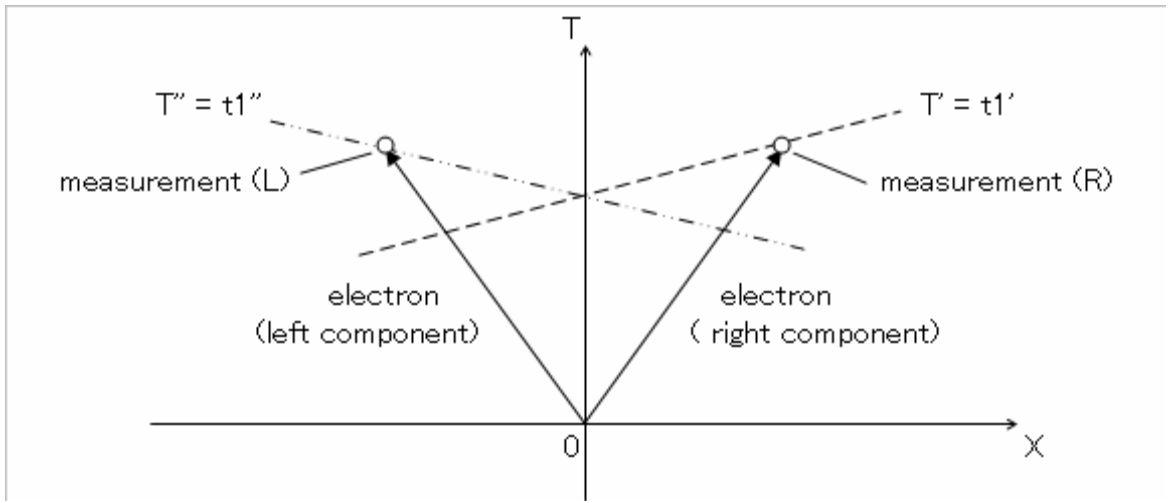

Fig. 2. Case1. Two measurement devices are going away from each other. T'= t1' line is the time when measurement (R) is done with respect to R. T"= t1" line is the time when measurement (L) is done with respect to L.

**4. Case2: A situation similar to violation of energy conservation law**

We consider the case in which two measurement devices are approaching to each other. In this case, as shown in Fig. 3, measurement (R) is done after measurement (L) with respect to R, and measurement (L) is done after measurement (R) with respect to L. According to above hypothesis, this means that each measurement is done before the collapse of wave packet caused by the other measurement. Therefore the result of each measurement must be independent from the other. So it is possible that both measurements detect electron. If amplitudes of right and left component of the electron are the same, the probability that both measurements detect the electron will be 1/4. Similarly, the probability that both measurements do not detect the electron will be 1/4. Fig. 3 shows the case that both measurements detect electron. This situation is similar to violation of energy conservation law because number of electron is not conserved.

So far we consider case2 using the concept of the collapse of wave packet. But under the many-worlds interpretation, at the time t2' when measurement (R) detect electron with respect to R, the correlation between the world in which measurement (R) detected electron and the world in which measurement (L) detected electron will vanish. Therefore if many-worlds interpretation is true, we can not observe above situation.

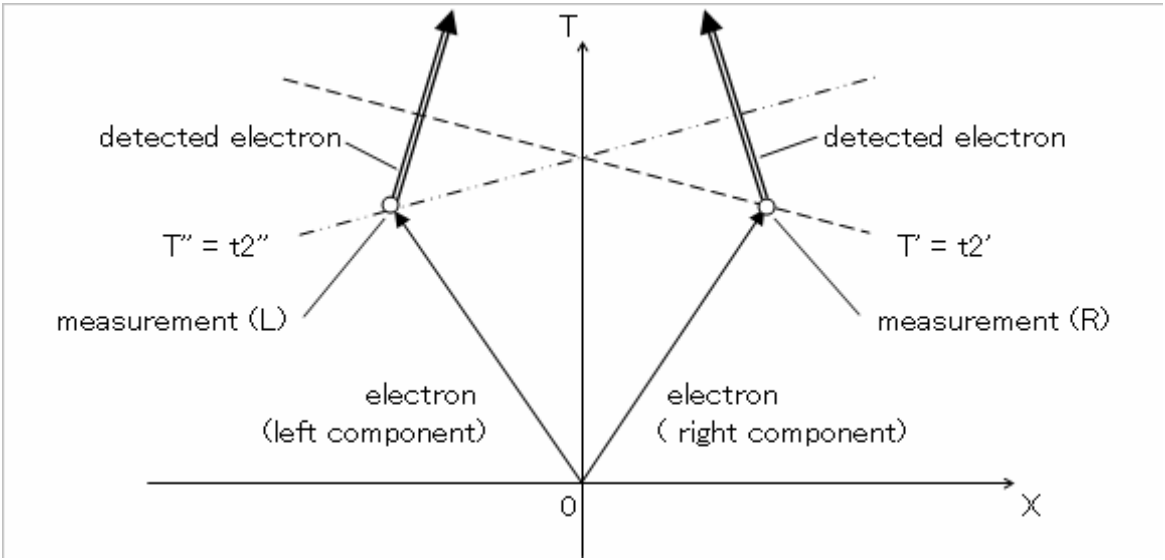

Fig. 3. Case2. Two measurement devices are approaching to each other. T'=t2' line is the time when measurement (R) is done with respect to R. T"=t2" line is the time when measurement (L) is done with respect to L.

## 5. Measurement devices moving mutually

So far we discussed about measurements by two measurement devices moving mutually. As shown in Fig. 4, these situations will be realized by using light or atomic beam which is projected on each component of split electron as the measurement devices in Case1 or Case2. And it will be possible to realize above paradoxical situations in experiment.

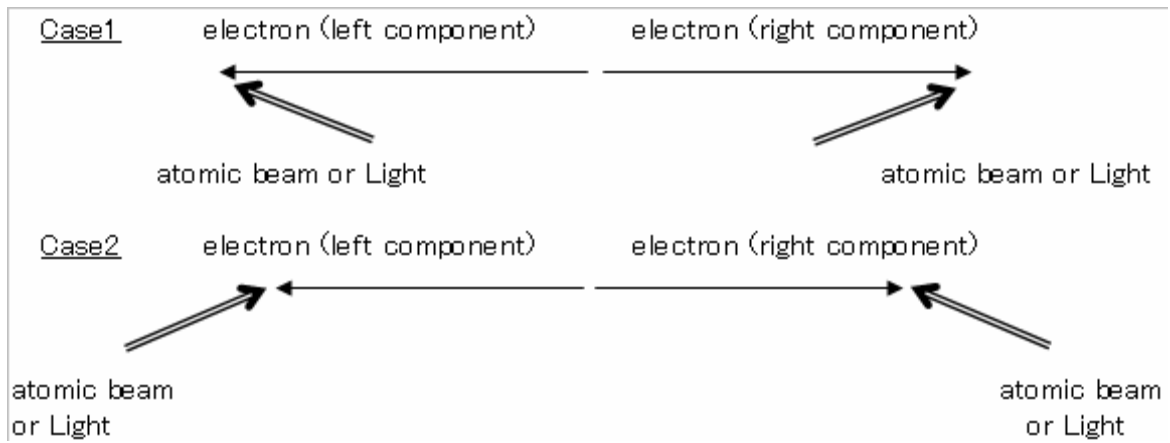

Fig. 4. Two atomic beams or light are projected on split electron at both sides

## 6. Conclusion

Entangled causal relation will be obtained in measurement by two measurement devices moving mutually. When two measurement devices are going away from each other, a situation in which the cause and the result can not be distinguished will be obtained. When two measurement devices are approaching to each other, a situation in which number of particle is not conserved will be obtained. These paradoxical results are derived from the concept of the collapse of wave packet and relativity of simultaneity without employing time machines.

# References

[1] M. S. Morris, K. S. Thorne and U. Yurtsever, Physical Review Letters, **61**, 1446-1449, (1988)

[2] J. R. Gott, Physical Review Letters, **66**, 1126-1129, (1991)

[3] F. Echeverria, G. Klinkhammer, and K. Thorne Physical Review D **44**, 1077-1099 (1991)

[4] D. Deutsch, Physical Review D **44**, 3197-3217 (1991)